\documentclass[a4paper,onecolumn,11pt,accepted=2023-06-05]{quantumarticle}
\pdfoutput=1
\usepackage[utf8]{inputenc}
\usepackage[english]{babel}
\usepackage[T1]{fontenc}

\usepackage{amsmath,amsthm,amssymb,amsfonts,mathrsfs}
\usepackage{dsfont}
\usepackage{mathabx}
\usepackage{centernot}
\usepackage{graphicx}
\usepackage{framed}
\usepackage{url}
\usepackage[dvipsnames]{xcolor}

\usepackage{hyperref}
\hypersetup{
	colorlinks = true,
	linkcolor = blue,
	citecolor = blue,
	urlcolor = blue,
	filecolor = black,
	linktoc = all,
}

\newcommand{\Tr}[1]{\operatorname{Tr}\!\left\{#1\right\}}
\newcommand{\PTr}[2]{\operatorname{Tr}_{{#1}}\!\left\{#2\right\}}

\def\id{\mathsf{id}}
\def\mE{\mathcal{E}}
\def\mK{\mathcal{K}}
\newcommand{\mI}{\mathcal{I}}
\newcommand{\mJ}{\mathcal{J}}
\newcommand{\mH}{\mathcal{H}}
\newcommand{\mIext}{\overline{\mI}}

\newcommand{\mHext}{\overline{\mH}}
\newcommand{\alice}{\textbf{\textrm{I}}}
\newcommand{\bob}{\textbf{\textrm{II}}}
\def\mD{\mathcal{D}}

\def\sH{\mathscr{H}}

\def\openone{\mathds{1}}
\newcommand{\set}[1]{\mathsf{#1}}
\newcommand{\op}[1]{\mathbb{#1}}

\newcommand{\vimplies}{\rotatebox[origin=c]{-90}{$\Leftarrow$}}
\newcommand{\dimplies}{\rotatebox[origin=c]{90}{$\Leftarrow$}}

\renewcommand{\ge}{\geqslant}

\def\>{\rangle}
\def\<{\langle}

\newtheorem{theorem}{Theorem} 

\newtheorem{corollary}{Corollary}
\newtheorem{definition}{Definition}

\theoremstyle{definition}

\theoremstyle{remark}
\newtheorem{remark}{Remark}

\newcommand{\bra}[1]{\langle #1 |}
\newcommand{\ket}[1]{|#1\rangle }
\newcommand{\ketbra}[1]{|#1\rangle \langle #1 |}

\begin{document}

\title{Unifying different notions of quantum incompatibility into a strict hierarchy of resource theories of communication}

\author{Francesco Buscemi}
\email{buscemi@i.nagoya-u.ac.jp}
\affiliation{Department of Mathematical Informatics, Nagoya University, Furo-cho, Chikusa-ku, 464-8601 Nagoya, Japan}
\orcid{0000-0001-9741-0628}
\author{Kodai Kobayashi}
\affiliation{Department of Mathematical Informatics, Nagoya University, Furo-cho, Chikusa-ku, 464-8601 Nagoya, Japan}
\author{Shintaro Minagawa}
\affiliation{Department of Mathematical Informatics, Nagoya University, Furo-cho, Chikusa-ku, 464-8601 Nagoya, Japan}
\orcid{0000-0002-8637-629X}
\author{Paolo Perinotti}
\affiliation{QUIT Group, Department of Physics, University of Pavia, via Bassi 6, 27100 Pavia, Italy}
\affiliation{INFN Sezione di Pavia, via Bassi 6, 27100 Pavia, Italy}
\orcid{ 0000-0003-4825-4264}
\author{Alessandro Tosini}
\affiliation{QUIT Group, Department of Physics, University of Pavia, via Bassi 6, 27100 Pavia, Italy}
\affiliation{INFN Sezione di Pavia, via Bassi 6, 27100 Pavia, Italy}
\orcid{0000-0001-8599-4427}
\maketitle
%
%
%
%

\begin{abstract}
    While there is general consensus on the definition of incompatible POVMs, moving up to the level of instruments one finds a much less clear situation, with mathematically different and logically independent definitions of incompatibility. Here we close this gap by introducing the notion of \emph{q-compatibility}, which unifies different notions of POVMs, channels, and instruments incompatibility into one hierarchy of resource theories of communication between separated parties. The resource theories that we obtain are \emph{complete}, in the sense that they contain complete families of free operations and monotones providing necessary and sufficient conditions for the existence of a transformation. Furthermore, our framework is fully \emph{operational}, in the sense that free transformations are characterized explicitly, in terms of local operations aided by causally-constrained directed classical communication, and all monotones possess a game-theoretic interpretation making them experimentally measurable in principle. We are thus able to pinpoint exactly what each notion of incompatibility consists of, in terms of information-theoretic resources.
\end{abstract}


\section{Introduction}

In the classical theory, measurements play no role: the theory assumes that all physical quantities can be measured simultaneously, accurately, and without disturbance. One of the most important new concepts brought about by quantum mechanics is that there now exist measurements that exclude each other: the mere act of performing one measurement unavoidably causes a disturbance that renders another one impossible. This was the intuition that Heisenberg had in mind when writing about his famous $\gamma$-ray microscope thought-experiment~\cite{heisenberg1927uber}. While Heisenberg's analysis is considered to be valid only on a heuristic level and under various (nowadays obsolete) assumptions~\cite{Ozawa-2015-Heisenberg-repeat}, modern and rigorous formulations of what is generally known as the ``uncertainty principle'' have been given in terms of preparation uncertainty relations~\cite{robertson,bialynicki1975uncertainty,maassen1988generalized,berta2010uncertainty}, measurement (i.e., noise--disturbance) uncertainty relations~\cite{ozawa-2003-universal-heisenberg,OZAWA-2004-uncertainty-generalized-measurements-annals,buscemi2008global,busch-lahti-werner,buscemi2014noise,ozawa-2019-soundness-completeness}, and incompatibility~\cite{busch-lahti-mittelstaedt,heinosaari2008notes,heinosaari2016invitation}. This work focuses on the latter.

While incompatibility is well understood and agreed-upon in the case in which only the outcomes statistics is considered (i.e., the case of POVMs), in the more general case, in which also the post-measurement state is included in the description of the measurement (i.e., for instruments), the situation is less clear. The main candidates for a notion of compatibility for instruments in the literature are classical compatibility~\cite{heinosaari-2014-strong-incompatibility}, also called traditional compatibility in~\cite{mitra2022compatibility}, and parallel compatibility~\cite{heinosaari2016invitation}. Both definitions have been advocated by different authors to provide the ``correct'' extension of the notion of compatibility to instruments: for example, while Ref.~\cite{ji2021incompatibility} argues in favor of classical compatibility, Ref.~\cite{mitra2022compatibility} instead argues in favor of parallel compatibility. The problem is that classical compatibility and parallel compatibility are logically inequivalent, as it is easy to construct counterexamples of measurements that are classically compatible but parallelly \textit{in}compatible, and vice versa. Hence, the question is whether a natural notion of compatibility for families of instruments exists, that fills the gap between classical and parallel compatibility.

In this work we propose two such notions: q-compatibility and no-exclusivity. Our definitions are given in clear operational terms, and framed within the general formalism of quantum resource theories~\cite{Chitambar-Gour2019resource-theories}. In particular, we show that all notions of incompatibility for instruments can be unified into a strict hierarchy of resource theories of constrained classical communication between two separated agents, similarly to what happens with entanglement and one-way or two-way LOCC transformations~\cite{horodecki-2009-review}. The resource-theoretic framework that we construct is \textit{complete}, in the sense that it contains a complete class of monotones and free operations, and \textit{operational}, in the sense that we explicitly characterize free operations as physical devices, avoiding purely convex-analytic concepts like resource morphisms~\cite{Zhou_2020}, that is, in this case, ``compatibility preserving operations''. Further, we prove a Blackwell-like theorem~\cite{blackwell1953,buscemi-CMP-2012} stating the equivalence between two \textit{incompatibility preorders} for families of instruments: a first preorder, capturing the idea that more incompatibility allows for higher payoffs in a class of guessing games, and a second preorder, which characterizes the possibility of transforming an initial, more incompatible family of instruments into another, less incompatible one. As a result, besides unifying previous inequivalent proposals, we are also able to characterize in a precise sense the operational meaning of quantum incompatibility.

\section{Definitions and notation}

In this work we only consider finite-dimensional quantum systems, i.e., systems associated with finite-dimensional Hilbert spaces. Systems are named $A,B,A',B',\dots$ and labeled as such, while the associated Hilbert spaces are denoted as $\sH_A, \sH_{B},\sH_{A'},\sH_{B'}\dots$ The space of all linear operators on $\sH$ is denoted as $\set{L}(\sH)$, and within it we have the set of positive semidefinite operators $\set{P}(\sH)$ and positive semidefinite operators with unit trace (i.e., density matrices or states) $\set{S}(\sH)$.

A \textit{channel} from $\set{L}(\sH_A)$ to $\set{L}(\sH_B)$ is defined as a completely positive trace-preserving linear map $\mE$, which we simply denote as $\mE:A\to B$. An \textit{instrument} is a family, indexed by a finite set $\set{X}$, of completely positive linear maps $(\mI_x)_{x\in\set{X}}$, each mapping $\set{L}(\sH_A)$ into $\set{L}(\sH_B)$, and such that their sum $\mI:=\sum_{x\in\set{X}}\mI_x$ is trace-preserving. Instruments are used to model quantum measurement processes~\cite{ozawa1984quantum}: given an initial state $\rho\in\set{S}(\sH_A)$, the number $p(x):=\Tr{\mI_x(\rho)}$ is interpreted as the predicted probability of occurrence of the outcome $x$, while $[p(x)]^{-1}\mI_x(\rho)\in\set{S}(\sH_B)$ is the corresponding post-measurement state. A convenient way to represent instruments is by using an additional ``classical pointer'' system $X$, such that the instrument can be written as a channel
\[
\rho\longmapsto\sum_{x\in\set{X}}\mI_x(\rho)\otimes\ketbra{x}_X\;,
\]
where $\<x'|x\>=\delta_{x,x'}$. In what follows, where needed, we will denote the above extended channel by adding a bar over the corresponding instrument, that is,
\[
\overline{\mI}(\bullet):=\sum_{x\in\set{X}}\mI_x(\bullet)\otimes\ketbra{x}_X\;.
\]
Notice that, in the above formula, while the input state is on $\sH_A$, the output state is bipartite on $\sH_B\otimes\sH_X$. To any instrument $(\mI_x)_{x\in\set{X}}$, there corresponds a unique \textit{POVM} (i.e., positive operator-valued measure), with elements given by $P_x:=\mI_x^\dag(\openone)\in\set{P}(\sH_A)$. In this expression, the map $\mI_x^\dag$ is the trace-dual of $\mI_x$, which is a map from $\set{L}(\sH_B)$ to $\set{L}(\sH_A)$. It is easy to verify that $\mI$ is trace-preserving if and only if $\sum_xP_x=\openone$, that is, if and only if the map $(\sum_x\mI_x^\dag)=\mI^\dag$ is unital (i.e., it maps the identity matrix $\openone_B$ to $\openone_A$).

In this paper we work in particular with \textit{families} of POVMs, channels, and instruments. A family of POVMs is denoted as
\[
(P_{x}^{(i)})_{x\in\set{X},i\in\set{I}}\;,
\]
where $i\in\set{I}$ is the index labeling the different POVMs, while $x\in\set{X}$ is the outcome index. For notational simplicity, even if different POVMs may have different outcome sets, we assume that we take an outcome set $\set{X}$ that is large enough to label the outcomes of all the POVMs in the family. In the case of families of channels we write
\[
(\mE^{(i)})_{i\in\set{I}}\;,
\]
where each $\mE^{(i)}$ is a channel from system $A_i$ to system $B_i$. We will often restrict to the case $A_i\equiv A$, for all $i$, however, if necessary, we will use the more detailed notation $(\mE^{(i)}:A_i\to B_i)_{i\in\set{I}}$. In the case of instruments, we denote a family as
\[
(\mI_{x}^{(i)})_{x\in\set{X},i\in\set{I}}\;,
\]
or $(\mI_{x}^{(i)}:A_i\to B_i)_{x\in\set{X},i\in\set{I}}$, when it is necessary to specify the input and output systems.

\section{A hierarchy of compatibilities}

In this paper we discuss four notions of compatibility for general families of quantum instruments, including families of POVMs and channels as special cases. Such notions are
\begin{enumerate}
    \item classical compatibility,
    \item parallel compatibility,
    \item q-compatibility, and
    \item no-exclusivity.
\end{enumerate}

\emph{Classical compatibility} constitutes the straightforward extension of the idea of POVMs compatibility~\cite{busch-lahti-mittelstaedt,heinosaari2008notes} to the case of instruments~\cite{heinosaari-2014-strong-incompatibility}. Classical compatibility, which is called ``traditional'' compatibility in Ref.~\cite{mitra2022compatibility}, is also the notion discussed in Ref.~\cite{ji2021incompatibility}. Here we use the word ``classical'' because this notion of compatibility can be understood entirely in terms of classical post-processings of the measurement's outcomes. \emph{Parallel compatibility}~\cite{heinosaari2016invitation} provides an alternative to classical compatibility. According to its definition, two or more instruments are said to be compatible if they can be simultaneously broadcast to separate receivers.

The main motivation for the present paper is that, as noticed in Ref.~\cite{mitra2022compatibility}, classical compatibility and parallel compatibility alone are not able to capture the whole picture of instruments compatibility, for it is not difficult to construct counterexamples of instruments that are classically compatible but parallelly incompatible, and vice versa. Thus, in a sense, both classical compatibility and parallel compatibility appear to be unnecessarily rigid.

The solution that we propose here takes as a basis the notion of \emph{q-compatibility}: a new, operationally motivated notion that contains both classical and parallel compatibilities as special cases. This means that now, given a family of instruments that is classically compatible but not parallelly compatible (or vice versa), one need not argue in favor of either definition to decide whether the given family is compatible or not, but simply treat it as a q-compatible family.

Q-compatibility, and thus classical and parallel compatibilities, are closely related to the intuitive notion of ``simultaneous measurability'', in the sense that they are oblivious of the order in which the measurements are given: they are defined as properties of a \textit{family} of measurements as a whole. On the contrary, the property of \emph{no-exclusivity,} recently introduced for pairs of instruments in Ref.~\cite{dariano2022incompatibility} and extended here to general families of instruments, depends on the order in which the instruments are performed: it is therefore not a property of a family of measurements, but of an \textit{ordered sequence} of measurements. In what follows, we will carefully introduce and analyze all these kinds of compatibilities, and derive the logical relations between them, as summarized in Table~\ref{table}.

\begin{table}[t]
    \centering
    \begin{tabular}{c|c}
      \hline 
       & notions of compatibility \\
      \hline \\
      instruments & $\left. \begin{array}{c} \text{classical compatibility}\\ \diagup\!\!\!\!\!\Uparrow \diagup\!\!\!\!\!\Downarrow\\ \text{parallel compatibility} \end{array} \right\} \Rightarrow \text{q-compatibility} \Rightarrow \text{no-exclusivity}$ \\ \\
      \hline \\
      channels & $\left. \begin{array}{c} \text{classical compatibility}\\ \diagup\!\!\!\!\!\Uparrow \diagup\!\!\!\!\!\Downarrow\\ \text{parallel compatibility} \end{array} \right\} \Rightarrow \text{q-compatibility} \Leftrightarrow \text{no-exclusivity}$ \\ \\
       \hline \\
      POVMs & $\text{classical compatibility} \Leftrightarrow \text{parallel compatibility} \Leftrightarrow \text{q-compatibility}$ \\ \\
      \hline
    \end{tabular}
            \caption{\label{table}A summary of the relations between different notions of compatibility, in the case of (from top to bottom): families of instruments, families of channels (i.e., instruments with singleton outcome set), and families of POVMs (i.e., instruments with trivial quantum output). See Theorem~\ref{thm:logical-rel} in the main text for the formal statement.}
\end{table}

\subsection{Classical compatibility}

\begin{figure}
        \centering
        \includegraphics[width=\textwidth,height=3.5cm,keepaspectratio]{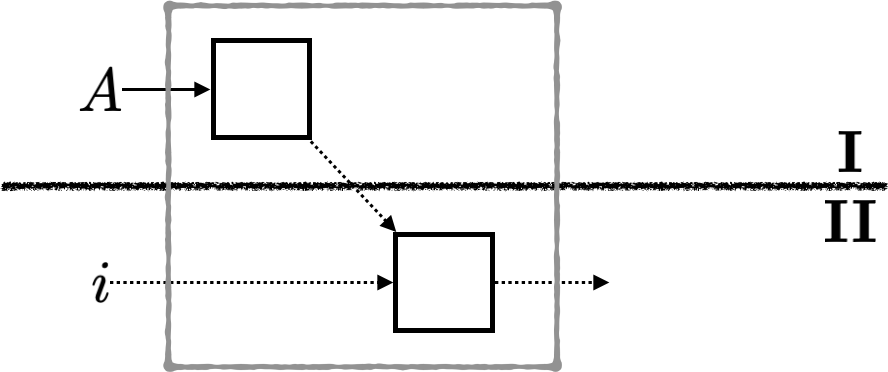}
        \caption{A programmable POVM (the gray outer box) that, by construction, can only provide a family of \textbf{compatible POVMs}~\cite{buscemi2020complete}. The converse also holds: any compatible family of POVMs can be obtained as a programmable POVM like the one depicted. Continuous arrows correspond to quantum communication, dotted lines to classical communication. The input index $i$ specifies the program, namely the instructions about which POVM should be implemented on system $A$.}
        \label{fig:comp-POVM}
\end{figure}

When looking only at the classical outcomes of measurement processes, that is, POVMs, the following notion of compatibility (which we therefore call \textit{classical} compatibility) is nowadays well established in the literature~\cite{heinosaari2008notes}:

\begin{framed}
\begin{definition}[compatibility for POVMs]\label{def:compatible-POVMs}
Given a family $(P_{x}^{(i)})_{x\in\set{X}, i\in\set{I}}$ of POVMs on system $A$, we say that the family is \emph{(classically) compatible}, whenever there exists 
\begin{itemize}
\item a ``mother'' POVM $(O_w)_{w\in\set{W}}$ on system $A$;
\item a family of conditional probability distributions $\mu(x|w,i)$,
\end{itemize}
such that
\begin{align}
    P_{x}^{(i)}=\sum_{w}\mu(x|w,i)O_w\;,
\end{align}
for all $x\in\set{X}$ and all $i\in\set{I}$. The situation is depicted in Fig.~\ref{fig:comp-POVM}.
\end{definition}
\end{framed}

\begin{remark}
Since the classical pointer state can be duplicated exactly, without loss of generality the POVM $(O_w)_w$ can be assumed to take outcomes in the product set $\set{X}_1\times\set{X}_2\times\cdots$, so that each $P_{x}^{(i)}$ can be obtained by simple marginalization, i.e., $P_{x}^{(i)}=\sum_{x_{i'}: i'\neq i}O_{x_1,x_2\dots}$. In fact, this marginalized form is the one usually found in the literature, especially in the case of families comprising two POVMs.
\end{remark}

\begin{remark}
Definition~\ref{def:compatible-POVMs} corresponds to the notion of full (simultaneous) compatibility. However, it is known that other weaker notions are possible: e.g., full compatibility is strictly stronger than pairwise compatibility~\cite{heinosaari2008notes}.
\end{remark}

A family of POVMs can always be thought of as a \textit{programmable} POVM, that is, a device with one classical output (i.e., the measurement outcome), and two inputs:
one quantum (i.e., the system's state being measured) and the other classical (i.e., the program deciding which POVM to measure). As argued in Ref.~\cite{buscemi2020complete}, the setup of a programmable device can be well understood if it is arranged in such a way that the two inputs are distributed between two separated parties: {\alice}, who receives the quantum input, and {\bob}---an inherently classical agent---who receives the classical program. The outcome, which is also classical information, is assumed to be with {\bob}. The case of compatible POVMs, that is, the case of programmable devices that can only implement compatible POVMs, corresponds to the situation, in which classical communication can only proceed from {\alice} to {\bob}, as schematically depicted in Fig.~\ref{fig:comp-POVM}. Essentially, {\alice}'s measurement constitutes the mother POVM, whose outcomes are sent to {\bob}, who in turns applies the correct post-processing depending of the input received. Any such a scheme necessarily leads to the implementation of compatible POVMs and, vice versa, any family of compatible POVMs can be implemented in such a scheme.

\begin{figure}
        \centering
        \includegraphics[width=\textwidth,height=3.5cm,keepaspectratio]{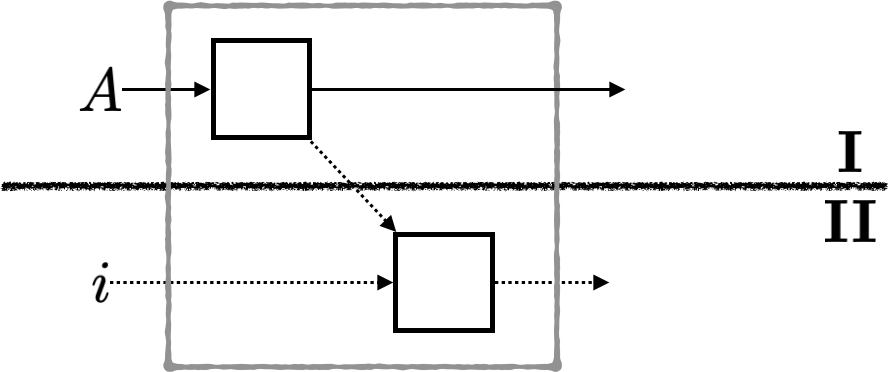}
        \caption{A programmable instrument (the gray outer box) that, by construction, can only provide a family of \textbf{classically compatible instruments}. The converse also holds: any classically compatible family of instruments can be obtained as a programmable instrument like the one depicted. Continuous lines correspond to quantum systems, dotted lines to classical information. The input index $i$ specifies the program.}
        \label{fig:class-comp-instr}
\end{figure}

Definition~\ref{def:compatible-POVMs} above can be directly extended from families of POVMs to families of instruments, thus obtaining the simplest and most stringent notion of compatibility for instruments as follows~\cite{heinosaari-2014-strong-incompatibility}:

\begin{framed}
\begin{definition}[classical compatibility for instruments]\label{def:class-compat-instr}
Given a family of instruments $(\mI_{x}^{(i)})_{x\in\set{X},i\in\set{I}}$, all mapping system $A$ to system $B$, we say that the family is \emph{classically compatible}, whenever there exists
\begin{itemize}
\item a mother instrument $(\mH_w)_{w\in\set{W}}$ from $A$ to $B$;
\item a family of conditional probability distributions $\mu(x|w,i)$,
\end{itemize}
such that
\begin{align}
    \mI_{x}^{(i)}=\sum_{w}\mu(x|w,i)\mH_w\;,
\end{align}
for all $x\in\set{X}$ and all $i\in\set{I}$. The situation is depicted in Fig.~\ref{fig:class-comp-instr}.
\end{definition}
\end{framed}

From the above definition and the corresponding Fig.~\ref{fig:class-comp-instr}, it is clear that, by construction, {\bob} cannot signal to {\alice}: the average channels $\mI^{(i)}=\sum_x\mI_{x}^{(i)}$ all coincide with $\mH=\sum_w\mH_w$. Indeed, as noticed in Ref.~\cite{ji2021incompatibility}, all classically compatible programmable instruments are necessarily {\bob}$\to${\alice} non-signaling. This also restricts the instruments in the family to have all the same output system $B$. Ref.~\cite{ji2021incompatibility} further explores the connection between classically compatible programmable instruments and the theory of non-signaling bipartite channels, including the steering scenario. In this work we depart from Ref.~\cite{ji2021incompatibility} in that we also consider signaling programmable instruments.

As it happens in the case of POVMs, it is possible to assume without loss of generality that the mother instrument already outputs a multi-index $(x_1,x_2,\dots)\in\set X\times \set{X}\times\dots$, so that each $\mI_{x}^{(i)}$ can be obtained by simple marginalization, i.e., $\mI_{x}^{(i)}=\sum_{x_{i'}: i'\neq i}\mH_{x_1,x_2\dots}$. In terms of the extended channel notation, each channel $\mIext^{(i)}:A\to BX$ can be obtained from $\mHext:A\to BX_1X_2\cdots$ by discarding all but one pointer systems, that is,
\begin{align}\label{eq:classical-marginalization}
    \mIext^{(i)}=\operatorname{Tr}_{X_{i'}:i'\neq i}\circ\;\mHext\;.
\end{align}
The above clearly constitutes a special case of channel post-processing~\cite{Buscemi-ProbInfTrans-2016,buscemi2018reverse-data-proc}.   Weaker, more general notions of instruments compatibility are obtained by allowing for more general post-processings on the mother instrument. In this regard, one should notice that simply appending an extra quantum post-processing box on {\alice}'s end before returning the quantum output would not add to the generality of the scheme, as such a post-processing could always be absorbed back into the mother instrument. As we will see in what follows, the way to introduce more general post-processings is to allow a further round of classical communication.

\subsection{Q-compatibility}

\begin{figure}
        \centering
        \includegraphics[width=\textwidth,height=3.5cm,keepaspectratio]{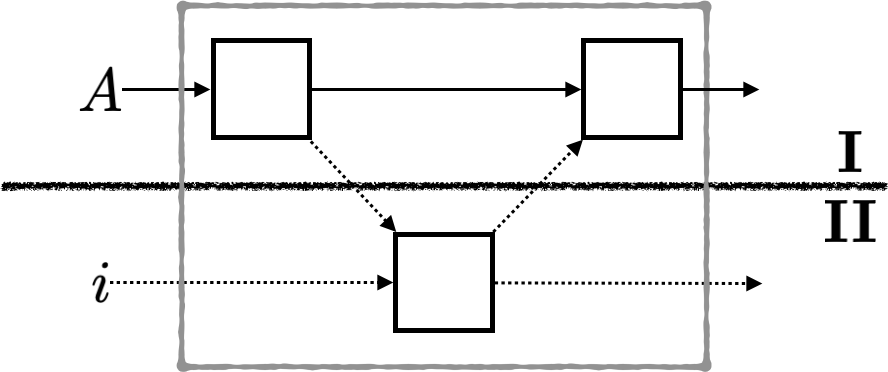}
        \caption{A programmable instrument (the gray outer box) that, by construction, can only provide a family of \textbf{q-compatible instruments}.  The converse also holds: any family of q-compatible instruments can be obtained as a programmable instrument like the one depicted. Continuous lines correspond to quantum systems, dotted lines to classical information. Note that quantum communication includes classical communication as a special case: hence, continuous lines can be though of as carrying both quantum and classical information simultaneously. The input index $i$ specifies the program.}
        \label{fig:strong-comp-instr}
    \end{figure}

While for POVMs only classical post-processing is possible, in the case of instruments, which also possess a quantum output, we can consider notions of compatibility that are strictly more general than classical compatibility (though they all fall back on classical compatibility in the special case of instruments with a trivial output systems, i.e., POVMs). In analogy with classical compatibility, which comes from the idea of post-processing the classical output, a generalized notion of compatibility arises when a post-processing is also performed on the quantum output. More precisely, we have the following:

\begin{framed}
\begin{definition}[q-compatibility]\label{def:strong-compat-instr}
Given a family of instruments $(\mI_{x}^{(i)})_{x\in\set{X},i\in\set{I}}$, all acting on the same system $A$ but with possibly different output systems $B_i$, we say that the family is \emph{q-compatible}, whenever there exist
\begin{itemize}
    \item a mother instrument $(\mH_w)_{w\in\set{W}}$ from $A$ to $C$;
    \item a family of conditional probability distributions $\mu(x|w,i)$;
    \item a family of channels $(\mD^{(x,w,i)}:C\to B_i)_{x\in\set{X},w\in\set{W},i\in\set{I}}$,
\end{itemize}
such that
\begin{align}
    \mI_{x}^{(i)}=\sum_{w}\mu(x|w,i)[\mD^{(x,w,i)}\circ\mH_w]\;,
\end{align}
for all $x\in\set{X}$ and all $i\in\set{I}$. The situation is depicted in Fig.~\ref{fig:strong-comp-instr}.
\end{definition}
\end{framed}

\begin{remark}
Notice how the post-processing channels may depend not only on $w$, but also on $x$ and $i$. This is possible because the notion of q-compatibility permits one more round of classical communication from {\bob} to {\alice}.
\end{remark}

Definition~\ref{def:strong-compat-instr} relaxes Definition~\ref{def:class-compat-instr} in various ways. For example, while Definition~\ref{def:class-compat-instr} can only be applied to families of instruments with the same input and output systems, in Definition~\ref{def:strong-compat-instr} it is possible to consider also the case of instruments with different output systems. More generally, Definition~\ref{def:strong-compat-instr} can deal with programmable instruments that are {\bob}$\to${\alice} signaling. For example, according to Definition~\ref{def:strong-compat-instr}, \textit{any family of channels $(\mE^{(i)})_i$ is automatically q-compatible} by construction. One can choose, for example, $\set{W}$ as the singleton, $\mH=\id$, and the post-processing channels $\mD^{(i)}\equiv\mE^{(i)}$ as the channels to be programmed. However, such a family \textit{cannot} be classically compatible, because it would allow signaling from {\bob} to {\alice}, as long as the channels $\mE^{(i)}$ are not all equal. That is to say, a family of channels is classically compatible if and only if all channels in the family coincide. The presence of signaling, however, does not constitute a problem for q-compatibility, as q-compatible programmable instruments have one round of classical communication going from {\bob} to {\alice} by construction. In other words, classical compatibility is \textit{strictly stronger} than q-compatibility: instruments that are not classically compatible, may still be q-compatible, but not vice versa.

\subsubsection{Parallel compatibility, no-broadcasting, and ``no information without disturbance''}

Another notion of compatibility that is stricly stronger than q-compatibility is that of \textit{parallel} compatibility. This is formally given as follows~\cite{heinosaari2016invitation}:

\begin{framed}
\begin{definition}[parallel compatibility]\label{def:parallel-compat-instr}
Given a family of instruments $(\mI_{x}^{(i)})_{x\in\set{X},i\in\set{I}}$, all acting on the same system $A$ but with possibly different output systems $B_i$, we say that the family is \emph{parallelly compatible}, whenever there exist
\begin{itemize}
    \item a mother instrument $(\mH_w)_{w\in\set{W}}$ from $A$ to $\otimes_{i\in\set{I}}B_i$;
    \item a family of conditional probability distributions $\mu(x|w,i)$,
\end{itemize}
such that
\begin{align}
    \mI_{x}^{(i)}=\sum_{w}\mu(x|w,i)[\operatorname{Tr}_{B_{i':i'\neq i}}\circ\;\mH_w]\;,
\end{align}
for all $x\in\set{X}$ and all $i\in\set{I}$.
\end{definition}
\end{framed}

In other words, parallelly compatible instruments are q-compatible instruments for which the post-processing channels $\mD^{(x,w,i)}$ in Definition~\ref{def:strong-compat-instr} can be restricted to partial traces, that is, $\mD^{(x,w,i)}=\operatorname{Tr}_{B_{i':i'\neq i}}$ for all $w$ and $x$. Parallel compatibility is therefore \textit{strictly stronger} than q-compatibility: instruments that are parallelly compatible are automatically q-compatible, but not vice versa.

Parallel compatibility provides a quantum analogue of classical compatibility, as expressed in Eq.~\eqref{eq:classical-marginalization}: as the latter is obtained by marginalization of the classical output coming from the mother, the former are obtained by marginalization of the mother's quantum output. On this point, however, it is necessary to raise one important \textit{caveat}: while classical post-processings can always be seen as marginalizations of a single broadcast channel (because classical information can be freely copied), quantum post-processings cannot be seen as such in general. This is the content of the \textit{no-broadcasting theorem}~\cite{no-broadcast} in quantum theory. In this respect, hence, the notion of parallel compatibility appears to be mixing two \textit{distinct} features of quantum theory: incompatibility, on the one side, and no-broadcasting, on the other\footnote{Surely, in a theory in which all states were broadcastable, then all measurements would automatically be compatible. The vice versa, however, is not true: it is possible to construct toy model theories in which all measurements are compatible, but states are not broadcastable \cite{PlaPC}. For example, quantum theory with only one informationally complete POVM and all its post-processings. We thus conclude that no-broadcasting is in general logically strictly stronger than compatibility, in the sense that the former implies the latter, but not vice versa. This is in line with the fact that parallel compatibility is strictly stronger than q-compatibility.}. One of the merits of q-compatibility is to untangle the two concepts.

More precisely, the concept of q-incompatibility is consistent with the notion of incompatibility as a quantitative version of the ``no information without disturbance'' principle of quantum theory: two instruments are, generally speaking, ``incompatible'' if there is no way to measure one without disturbing the other. The notion of parallel incompatibility seems to point in the \textit{opposite} direction instead: noiseless instruments \textit{cannot possibly ever} be parallelly compatible, even though by definition they cause no disturbance. In fact, the less disturbing two instruments are, the less parallelly compatible they can be. For example, according to the definition of parallel compatibility, the identity channel is \textit{not} compatible with itself, because that would amount to perfect broadcasting, which contradicts the no-broadcasting theorem.

Therefore, even though parallel compatibility seems like a natural \textit{analogue} of classical compatibility (as they both arise from suitable marginalizations), parallel compatibility does not constitutes an \textit{extension} thereof. In fact, the two notions are \textit{logically inequivalent}, since neither implies the other~\cite{mitra2022compatibility}. Hence, another advantage of the concept of q-compatibility introduced here is that it is able to close the gap, as now both classical and parallel compatibility become special cases of q-compatibility.

\subsection{No-exclusivity}

\begin{figure}
        \centering
        \includegraphics[width=\textwidth,height=3.5cm,keepaspectratio]{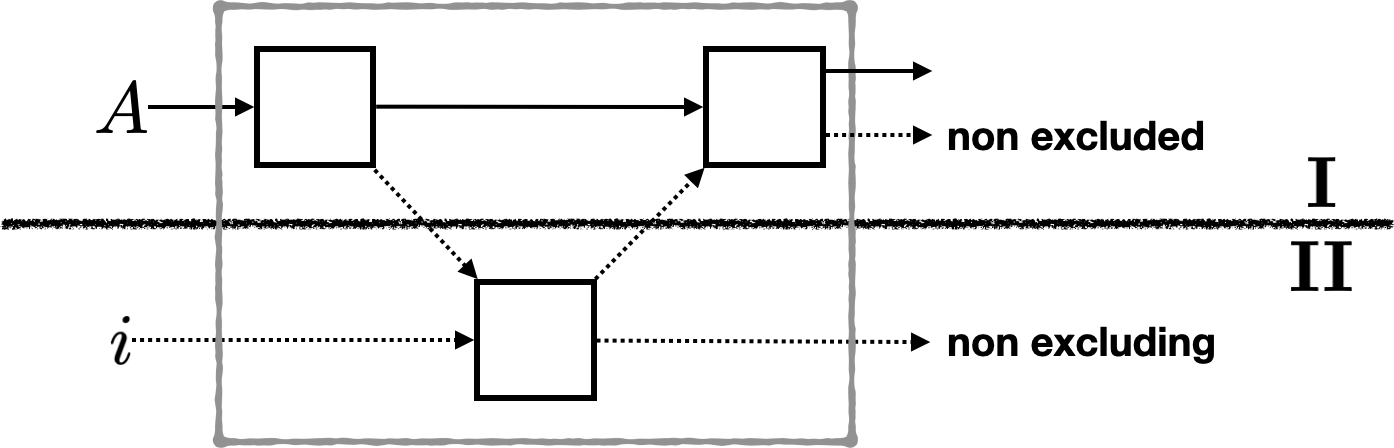}
        \caption{A programmable instrument (the gray outer box) that can provide {\bob} only with the outcomes of \textbf{non-exclusive instruments}. The converse also holds: if the programmable instrument is non-exclusive, it can always be obtained like the one depicted, where the outcomes held {\bob} correspond to non-exclusive instruments, while {\alice} holds the outcomes of the non-excluded ones. Continuous lines correspond to quantum systems, dotted lines to classical information. Note that quantum communication includes classical communication as a special case: hence, continuous lines can be though of as carrying both quantum and classical information simultaneously. The input index $i$ specifies the program.}
    \label{fig:weak-comp-instr}
\end{figure}

Recently Ref.~\cite{dariano2022incompatibility} proposed a further weakened notion of compatibility named \textit{no-exclusivity}. The idea is originally formulated for two instruments as follows: given two instruments $(\mI_{x})_{x\in\set{X}}$ and $(\mJ_{y})_{y\in\set{Y}}$, we say that $(\mI_{x})_{x\in\set{X}}$ does not exclude $(\mJ_{y})_{y\in\set{Y}}$ if it is possible to implement $(\mI_{x})_{x\in\set{X}}$ first, in such a way that at a later time it is possible to use its quantum output along with any quantum memory to implement $(\mJ_{y})_{y\in\set{Y}}$, in which case only the quantum output of the latter can be produced. More precisely:
\begin{framed}
\begin{definition}[no-exclusivity]
Given two instruments $(\mI_{x})_{x\in\set{X}}$ and $(\mJ_{y})_{y\in\set{Y}}$, both acting on the same system $A$ but with possibly different output systems $B_1$ and $B_2$, we say that $(\mI_x)_{x\in\set X}$ \emph{does not exclude} $(\mJ_y)_{y\in Y}$, if there exist
\begin{itemize}
    \item  a mother instrument $(\mH_{w})_{w\in\set W}$ from $A$ to $C$;
    \item a conditional probability distribution $\mu(x|w)$;
    \item a family of channels $\{\mD^{(x,w)}:C\to B_1\}_{x\in\set{X},w\in\set{W}}$;
    \item a family of instruments $(\mK^{(w)}_y)_{y\in\set Y,w\in\set w}$ from $C$ to $B_2$,
\end{itemize}
such that
\begin{align}
    &\mI_x=\sum_{w}\mu(x|w)[\mD^{(x,w)}\circ \mH_w]
    \;,\label{eq:no-excl}\\
    &\mJ_y=\sum_w\mK_{y}^{(w)}\circ \mH_w\nonumber\;,
\end{align} 
for all $x\in\set{X}$ and all $y\in\set{Y}$. The situation is depicted in Fig.~\ref{fig:weak-comp-instr}.
\end{definition}
\end{framed}
\begin{remark}
    The original definition of no-exclusivity given in Ref.~\cite{dariano2022incompatibility} only considers the partial trace as quantum post-processing channel, in analogy with the definition of parallel compatibility. Here we generalize it, so that it is immediate to recognize that no-exclusivity constitutes a further relaxation of the notion of q-compatibility.
\end{remark}

Notice that the classical outcome of the instrument done first can be retained, so that in the end both outcomes can be had. In other words, as one would expect, also the notion of no-exclusivity reduces to the notion of classical compatibility in the case of instruments with no quantum output. Notice moreover that, in general, the relation of no-exclusivity is \textit{not} symmetric: the fact that there exists a way to implement $(\mI_{x})_{x\in\set{X}}$ without excluding $(\mJ_{y})_{y\in\set{Y}}$, does not imply that also the vice versa holds---and even in that case, the mother instrument and the set of post-processings may differ.

Since the notion of no-exclusivity crucially depends on the order in which two instruments are measured, in the case of families of instruments comprising more than two elements, it is easier to express the idea of no-exclusivity in the negative form. We thus introduce the following definition for a \textit{family} of instruments:

\begin{framed}
	\begin{definition}[exclusivity]\label{def:exclusivity}
		Given a family of instruments $(\mI_{x}^{(i)})_{x\in\set{X},i\in\set{I}}$, all acting on the same system $A$ but with possibly different output systems $B_i$, we say that the family is \emph{exclusive}, whenever for any $i_0\in\set{I}$, there exists at least one among the remaining instruments $(\mI_{x}^{(i)})_{x\in\set{X},i\in\set{I}\setminus\{i_0\}}$ that is excluded by $(\mI_{x}^{(i_0)})_{x\in\set{X}}$.
	\end{definition}
\end{framed}

Accordingly, a family of instruments $(\mI_{x}^{(i)})_{x\in\set{X},i\in\set{I}}$ is said to be \textit{non-exclusive} if there exists an index $i_0\in\set{I}$ such that the corresponding instrument $(\mI_{x}^{(i_0)})_{x\in\set{X}}$ can be implemented without precluding the possibility of later implementing \textit{any one} among the remaining instruments $(\mI_{x}^{(i)})_{x\in\set{X},i\in\set{I}\setminus\{i_0\}}$.

\begin{remark}
	Notice the difference between the above definition and the following property, which we could call \textit{full exclusivity}: a family of instruments is fully exclusive whenever for any $i_0\in\set{I}$, all remaining instruments $(\mI_{x}^{(i)})_{x\in\set{X},i\in\set{I}\setminus\{i_0\}}$ are excluded by $(\mI_{x}^{(i_0)})_{x\in\set{X}}$. The difference between the two notions lies in the use of quantifiers. While exclusivity (Definition~\ref{def:exclusivity}) can be summarized as:
	\begin{equation*}
		\forall i_0:\exists i\in\set{I}\setminus\{i_0\}\text{ such that } (\mI^{(i)}_x)_{x\in\set{X}} \text{ is excluded,}
	\end{equation*}
	full exclusivity instead reads:
	\begin{equation*}
		\forall i_0: \forall i\in\set{I}\setminus\{i_0\}\;,\;  (\mI^{(i)}_x)_{x\in\set{X}} \text{ is excluded.}
	\end{equation*}
\end{remark}

Let us now assume that the given family $(\mI_{x}^{(i)})_{x\in\set{X},i\in\set{I}}$ is non-exclusive. In such a case, we are in a situation like the one depicted in Fig.~\ref{fig:weak-comp-instr}: there is an instrument, which is the one implemented beforehand by {\alice} and whose outcomes are with {\bob}, that by construction does not exclude any subsequent instrument done by {\alice}. The formulation of no-exclusivity thus requires a modification also in the way in which the bipartition {\alice}/{\bob} is understood: while in the case of classical and q-compatibilities we could neatly separate {\alice} and {\bob} depending on the nature of their outputs (quantum for {\alice}, classical for {\bob}), we now need to separate the outcomes into those (with {\bob}) corresponding to the non-exclusive instrument, and those (with {\alice}) corresponding to the non-\textit{excluded} instruments.

We summarize our discussion so far in the following statement (see also Table~\ref{table}): 

\begin{framed}
\begin{theorem}\label{thm:logical-rel}
The logical relations between the various notions of compatibility for \emph{families of general quantum instruments} are as follows
\[
\left.\begin{matrix}
	\text{classical compatibility}\\
	\diagup\!\!\!\!\!\!\vimplies\ \diagup\!\!\!\!\!\!\dimplies\ \\
	\text{parallel compatibility}
\end{matrix}\right\}
\implies\ \text{q-compatibility}\ \implies\ \text{no-exclusivity},
\]
and all implications are strict.

When restricted to \emph{families of channels}, that is, programmable instruments with singleton outcome set, we have
\[
\left.\begin{matrix}
	\text{classical compatibility}\\
	\diagup\!\!\!\!\!\!\vimplies\ \diagup\!\!\!\!\!\!\dimplies\ \\
	\text{parallel compatibility}
\end{matrix}\right\}
\implies\ \text{q-compatibility}\ \iff\ \text{no-exclusivity}.
\]
In particular, a family of channels is always (trivially) q-compatible, while it is classical compatible if and only if all channels in the family are the same channel.

When restricted to \emph{families of POVMs}, that is, programmable instruments with trivial quantum output (i.e., $B=\mathbb{C}$), the following equivalence holds
\[
\text{classical compatibility}\ \iff\ \text{parallel compatibility}\ \iff\ \text{q-compatibility}.
\]
\end{theorem}
\end{framed}

\begin{proof}
    The fact that classical compatibility and parallel compatibility are independent notions comes from the following counterexamples already mentioned in the main text. First, two channels can be parallelly compatible without being the same channel, and since classical compatibility implies {\bob}$\to${\alice} no-signaling, such pair is not classically compatible. Conversely, the family comprising two channels $(\id,\id)$ is classically compatible, but cannot be parallelly compatible due to the no-broadcasting theorem.

    Next, q-compatibility includes both classical compatibility and parallel compatibility because the class of post-processings allowed in the definition of q-compatibility includes both classical post-processings (and hence, classical compatibility) and partial traces (and hence, parallel compatibility). The inclusion is strict because a family of channels is always q-compatible even if the channels are neither different (and hence not classically compatible) nor broadcastable (and hence not parallelly compatible).

    Finally, no-exclusivity is a relaxation of q-compatibility, because in the former {\alice} can use instruments, and not just channels, for post-processing. The inclusion is strict because the family comprising three instruments $(\id,(\mI_x)_{x\in\set{X}},(\mJ_y)_{y\in\set{Y}})$, where $\id$ is the trivial (singleton, non-disturbing) instrument, and $(\mI_x)_{x\in\set{X}}$ and $(\mJ_y)_{y\in\set{Y}}$ are two q-incompatible instruments, is non-exclusive (because $\id$ does not exclude $(\mI_x)_x$ nor $(\mJ_y)_y$) but it is not q-compatible (because $(\mI_x)_x$ and $(\mJ_y)_y$ are not).

    The particular case of channels was discussed after Definition~\ref{def:strong-compat-instr}. Finally, the equivalence of classical compatibility, parallel compatibility, and q-compatibility in the case of families of POVMs can be easily understood, because {\bob} is always able to output the outcomes of all the instruments in the family.
\end{proof}

\section{A hierarchy of resource theories}

In this section we explicitly develop the actual resource-theoretic framework, in which the resources to be manipulated are programmable instruments. Operations are, therefore, \textit{superoperations}, i.e., supermaps~\cite{chiribella2008transforming} transforming programmable instruments into programmable instruments. Furthermore, suitable restrictions, in the form of limitations on how and what kind of information can be exchanged between {\alice} and {\bob}, are imposed on such superoperations.

\subsection{Free superoperations for classical incompatibility}

We begin from the most stringent scenario, in which only classically compatible programmable instruments are freely available. This scenario is the one considered also in Ref.~\cite{ji2021incompatibility}, but here we use a class of free operations different from that of~\cite{ji2021incompatibility}. This is because we want a class of superoperations that can be later extended to accommodate signaling programmable instruments, whereas the class of free operations considered in~\cite{ji2021incompatibility} can only work for non-signaling devices.

\begin{figure}
        \centering
        \includegraphics[width=\textwidth,height=3.5cm,keepaspectratio]{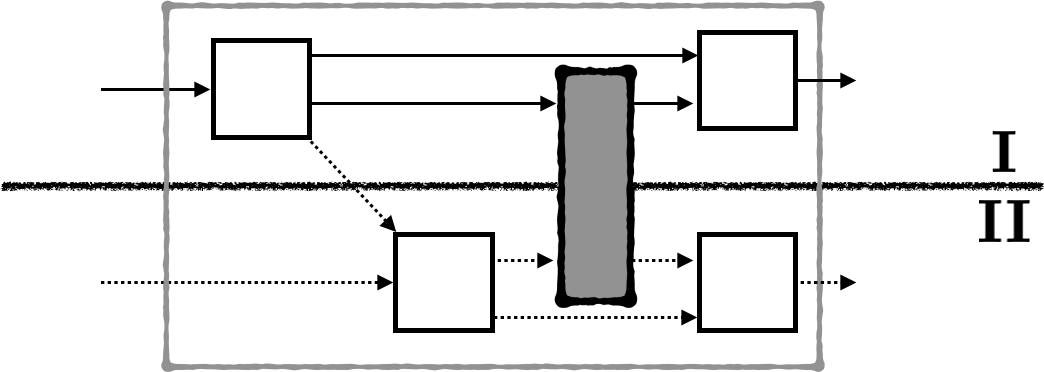}
        \caption{A complete class of free superoperations for \textbf{classically compatible} programmable instruments. If the programmable instrument provided (the dark gray inner box) is classically compatible, then the resulting light gray outer box is also classically compatible. Moreover, all classically compatible programmable instruments can be created for free and are thus equivalent to each other under the operational paradigm depicted. Continuous lines correspond to quantum systems, dotted lines to classical information. Note that quantum communication includes classical communication as a special case: hence, continuous lines can be though of as carrying both quantum and classical information simultaneously.}
        \label{fig:class-comp-free-op}
\end{figure}

Since the crucial aspect of classically compatible programmable instruments is their {\bob}$\to${\alice} non-signaling property (see Fig.~\ref{fig:class-comp-instr}), we assume that free superoperations can only use classical {\alice}$\to${\bob} communication. Beside this, the two agents are allowed to use any local side-resource of their choosing. We thus define free superoperations as follows: {\alice} can perform any pre-processing instrument, send the classical output to {\bob}, and use any amount of quantum memory on her side, while {\bob} is allowed to processes the classical part of the device using the classical information received from {\alice} and any noiseless classical side-channel. Within the context of instrument compatibility, we refer to such a class of one-way LOCC superoperations as \textit{classically compatible superoperations}. The situation is depicted in Fig.~\ref{fig:class-comp-free-op}.

Just by comparing Fig.~\ref{fig:class-comp-free-op} with Fig.~\ref{fig:class-comp-instr} it is easy to see that any classically compatible programmable instrument can be implemented for free in this scenario. Moreover, if {\alice} and {\bob} process a classically compatible programmable instrument using a classically compatible superoperation, they can only end up with another classically compatible programmable instrument. In order to see this, one can just substitute the dark gray box in Fig.~\ref{fig:class-comp-free-op} with a classically compatible programmable instrument as explicitly constructed in Fig.~\ref{fig:class-comp-instr}, and notice that the resulting device can again be compactly represented in the same way. In other words, classical communication keeps flowing only from {\alice} to {\bob}.

An important point to stress here, and that will apply to the following discussion as well, is that we are not claiming that this model of free superoperations is \textit{the only one} that have classically compatible programmable instruments as free resources: the same resource may arise from different operational scenarios. One example is provided by Ref.~\cite{ji2021incompatibility} in which another paradigm is adopted. A certain degree of freedom in the definition of the operational framework constitutes a general feature of resource theories: another example is provided by the resource theory of entanglement, in which separable states constitute the free resources both under the LOSR~\cite{buscemi2012all,schmid-LOSR} and the LOCC~\cite{horodecki-2009-review} paradigms, for example.

\subsection{Free superoperations for q-incompatibility}

\begin{figure}
        \centering
        \includegraphics[width=\textwidth,height=3.5cm,keepaspectratio]{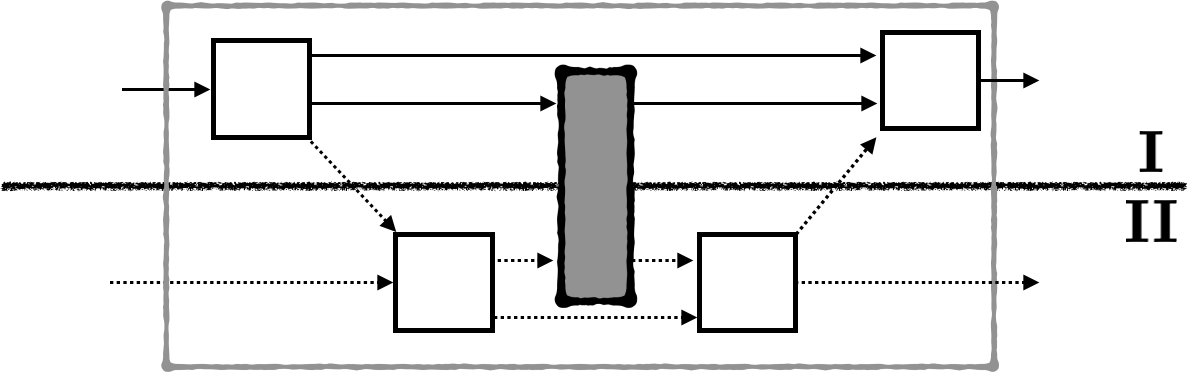}
        \caption{A complete class of free superoperations for \textbf{q-compatible} programmable instruments. If the programmable instrument provided (the dark gray inner box) is q-compatible, then the resulting light gray outer box is also q-compatible. Moreover, all q-compatible programmable instruments can be created for free and are thus equivalent to each other under the operational paradigm depicted. Continuous lines correspond to quantum systems, dotted lines to classical information. Note that quantum communication includes classical communication as a special case: hence, continuous lines can be though of as carrying both quantum and classical information simultaneously.}
        \label{fig:strong-comp-free-op}
    \end{figure}

In order to characterize a suitable set of free superoperations in a resource theory of q-in\-com\-pa\-ti\-bi\-li\-ty, we need to provide {\alice} and {\bob} with a wider set of allowed operations with respect to those allowed in the resource theory of classical incompatibility. More precisely, we now allow for one more round of classical communication from {\bob} to {\alice} after the dark gray box (i.e., the resource initially shared by the two players) has been used. The situation is depicted in Fig.~\ref{fig:strong-comp-free-op}. Again, just by comparing Fig.~\ref{fig:strong-comp-free-op} with Fig.~\ref{fig:strong-comp-instr}, it is easy to see that whenever the dark gray box is a q-compatible programmable instrument, the players, by exploiting the resources given to them, can only construct another q-compatible programmable instrument. This is due to the fact that, in the end, the final device has only one round of classical communication from {\alice} to {\bob}, followed by one round of classical communication back from {\bob} to {\alice}. There is no further interaction between the players. In particular---and most importantly---there is no further {\alice}$\to${\bob} signaling after {\alice} has received information from {\bob}. If this happens, then \textit{any} programmable instruments, even \textit{incompatible} ones, become free: indeed, in such a case {\alice} can send to {\bob} the classical output of a measurement she does locally, after she has learned the ``which measurement'' information, thus trivializing the framework.

Another possibility for the players to create incompatibility for free would be to substitute classical communication with quantum communication, by means of direct quantum communication or teleportation: in such a case, {\alice} could transmit the input state to {\bob}, who could perform the instrument locally after having learned the ``which measurement'' information, send the quantum output back to {\alice}, and thus render the whole framework trivial again. In any case, if the players can create incompatibility, then it necessarily means that they possess resources that go beyond the allowed ones: for example, one more round of classical communication, or bidirectional quantum communication, or free ebits to use in a teleportation protocol, etc. The vice versa is also true: as we will explicitly show in Section~\ref{sec:monotones} below, it is possible to construct a game-theoretic scenario in which two players possessing resources beyond the allowed ones would be able to consistently win strictly more often than expected in a whole class of suitable games.

We conclude this section with a comment on the resource theory of parallel incompatibility. Since the question of whether two channels are parallelly compatible coincides with the channels marginal problem \cite{PhysRevResearch.4.013249}, the resource theory of parallel incompatibility seems to be more related to a resource theory of unextendibility~\cite{PhysRevLett.123.070502,PhysRevA.104.022401} than to a resource of causally-constrained directed classical communication, such as those considered here. We leave this line of inquiry open for future investigation.
    
\subsection{Free superoperations for exclusivity}
 
     \begin{figure}
        \centering
        \includegraphics[width=\textwidth,height=3.5cm,keepaspectratio]{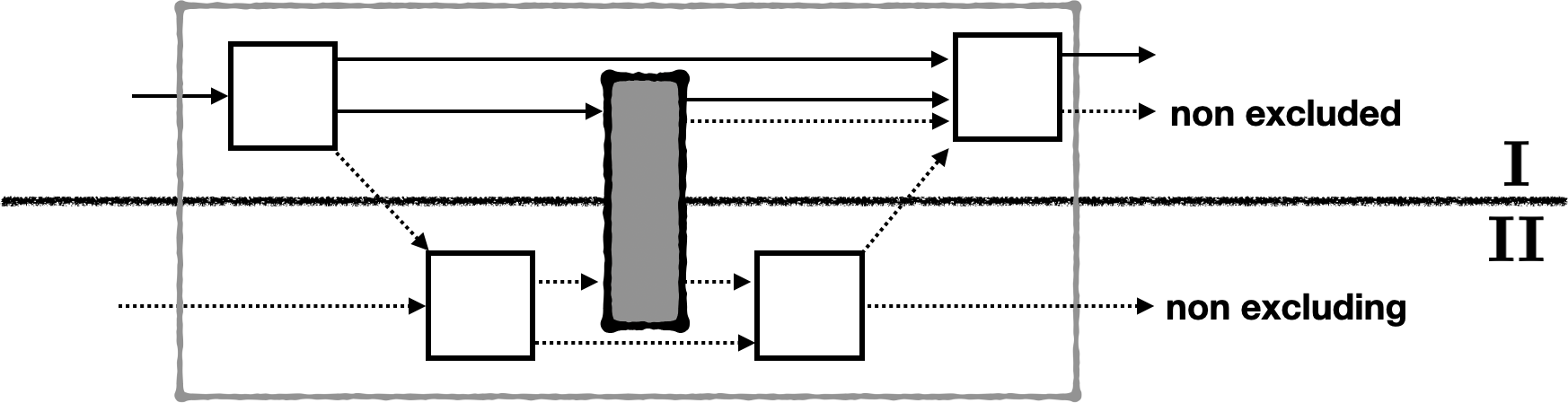}
        \caption{A complete class of free superoperations for \textbf{non-exclusive programmable} instruments. If the programmable instrument initially shared between {\alice} and {\bob} (the dark gray inner box) is able to provide {\bob} only with the outcomes of non-exclusive instruments, then also the resulting light gray outer box necessarily gives {\bob} only the outcomes of non-exclusive instruments. Moreover, all non-exclusive programmable instruments can be created for free and are thus equivalent to each other under the operational paradigm depicted. Continuous lines correspond to quantum systems, dotted lines to classical information. Note that quantum communication includes classical communication as a special case: hence, continuous lines can be though of as carrying both quantum and classical information simultaneously.}
        \label{fig:weak-comp-free-op}
    \end{figure}

We understand a resource theory of exclusivity as the possibility for {\bob} to output the outcomes of instruments that can exclude others in the family. Hence, a free programmable instrument, in the resource theory of exclusivity, is any programmable instrument such that {\bob}'s outcomes are all about non-exclusive measurements. Correspondingly, in order to capture operationally this idea, we further extend the players capabilities, by allowing Alice the possibility of finally performing post-processing \textit{instruments}, instead of just channels. In this way, {\alice} can locally produce not only the quantum outputs, but also the classical outcome of the other instruments, which can thus be quite general. The constraints are therefore weaker, with respect to the previous situations, in which {\bob} was required to output the outcomes of \textit{all} instruments in the family. See Fig.~\ref{fig:weak-comp-free-op} for a schematic diagram.

\section{Majorization preorders of programmable instruments}\label{sec:monotones}

In this section we develop the theory of statistical comparison for the resource theories of incompatibility that we introduced above. Statistical comparison is a concept introduced by Blackwell~\cite{blackwell1953} with the aim of extending the ideas of Lorenz curves and majorization~\cite{hardy1952inequalities,MarshallOlkin} to statistical objects more general than just pairs of probability distributions such as, e.g., statistical models or noisy channels. The seminal result, called the Blackwell--Sherman--Stein Theorem, proves the equivalence between two natural preorders in this scenario: one, which is analogous to the majorization preorder, and is given by the existence of a suitable transformation (e.g., a doubly stochastic matrix, in the case of majorization) between two objects; and another, which, analogously to the comparison of the Lorenz curves, is constructed instead by comparing the expected performance of some class of statistical tests (e.g., hypothesis testing, in the case of Lorenz curves) done on the objects at hand. Such a viewpoint, equating the existence of a transformation with the comparison of a class of operational utilities (in resource-theoretic jargon, \textit{monotones}), summarizes the core concept that lies at the basis of all resource theories~\cite{Chitambar-Gour2019resource-theories}. Indeed, recent quantum extensions of Blackwell's theory of statistical comparison~\cite{buscemi-CMP-2012,Buscemi-ProbInfTrans-2016,jencova-comparison-2015,buscemi-gour-2017-q-lorenz,buscemi2018reverse-data-proc,Buscemi-sutter-tomamichel-2019information,jencova-comparison-2021} have found fruitful applications in many specific resource-theoretic scenarios, including entanglement and nonlocality theory, information theory, open quantum systems dynamics, quantum coherence, and quantum thermodynamics~\cite{buscemi2012all,buscemi-datta-strelchuk-2014,buscemi-2015-fully-quantum-second-laws,buscemi-datta-2016-divisibility,Rosset-Buscemi-Liang-PRX,Gour-Jennings-Buscemi_2018,Skr-Linden-2019,regula-buscemi-2020-coherence-prr,buscemi2020complete,Schmid2020typeindependent,Zhou_2020,rosset-schmid-buscemi-2020-prl,ji2021incompatibility,buscemi2023sharpness}. The construction that we present here is closely related to that employed in Refs.~\cite{buscemi2020complete} and~\cite{ji2021incompatibility}, but it contains some adaptations to make it suitable to the present operational framework.

We begin by introducing three incompatibility majorization preorders, each corresponding to the notions of classical compatibility, q-compatibility, and non-exclusivity, respectively:
\begin{definition}[incompatibility majorization preorders]
Given two programmable instruments $(\mI_{x}^{(i)}:A\to B_i)_{x\in\set{X},i\in\set{I}}$ and $(\mJ_{y}^{(j)}:C\to D_j)_{y\in\set{Y},j\in\set{J}}$, we say that $(\mI_{x}^{(i)})_{x,i}$ is ``more classically incompatible'' than $(\mJ_{y}^{(j)})_{y,j}$, and write
\[
(\mI_{x}^{(i)})_{x,i}\succeq_c(\mJ_{y}^{(j)})_{y,j}\;,
\]
whenever there exists a classically compatible superoperation (see Fig.~\ref{fig:class-comp-free-op}) that is able to transform $(\mI_x^{(i)})_{x,i}$ into $(\mJ_y^{(j)})_{y,j}$.

Perfectly analogous definitions are given also for the ``more q-incompatible'' preorder $\succeq_q$ (where free superoperations are as in Fig.~\ref{fig:strong-comp-free-op}), and the ``more exclusive'' preorder $\succeq_{ex}$ (Fig.~\ref{fig:weak-comp-free-op}).
\end{definition}

In what follows, we provide the operational, game-theoretic formulation of the preorders introduced above. Let us consider two programmable instruments, $(\mI_{x}^{(i)}:A\to B_i)_{x\in\set{X},i\in\set{I}}$ and $(\mJ_{y}^{(j)}:C\to D_j)_{y\in\set{Y},j\in\set{J}}$. We want to reformulate, in an equivalent way, the condition
\begin{equation}\label{eq:condition}
\exists \textrm{ superoperation } \op{T}:[\op{T}\mI]_{y}^{(j)}=\mJ_{y}^{(j)},\quad\forall j\in\set{J},y\in\set{Y}\;.
\end{equation}
In the above, the transformation $\op{T}$ denotes a superoperation that belongs to the class of classically compatible superoperations (denoted by $\set{T}_c$), q-compatible superoperations (denoted by $\set{T}_q$), or non-exclusive superoperations (denoted by $\set{T}_{ex}$), respectively, corresponding to the preorder, $\succeq_c$, $\succeq_q$, or $\succeq_{ex}$, respectively. The juxtaposition $\op{T}\mI$ denotes the programmable instrument obtained by processing (via the superoperation $\op{T}$) the initial resource $(\mI_{x}^{(i)})_{x,i}$. In Figs.~\ref{fig:class-comp-free-op},  \ref{fig:strong-comp-free-op}, and~\ref{fig:weak-comp-free-op}, the initial programmable instrument $(\mI_{x}^{(i)})_{x,i}$ is represented as the inner dark gray box, while the resulting programmable instrument $([\op{T}\mI]_{y}^{(j)})_{y,j}$ is represented as the outer light gray box. Notice that the resulting programmable instrument $\op{T}\mI$ is assumed to be structurally consistent with the ideal target $(\mJ_{y}^{(j)})_{y,j}$, sharing the same program set, the same outcome set, the same input systems, and the same output systems.

The notation used in Eq.~\eqref{eq:condition} is therefore rather implicit, but it allows us to consider all preorders at once. The important point is that the three sets of free superoperations, $\set{T}_{c}$, $\set{T}_{q}$, and $\set{T}_{ex}$ are all closed and convex. This fact is a consequence of the classical communication from {\alice} to {\bob} that is present is all frameworks: indeed, at the beginning of the transformation protocol, {\alice} can locally randomize her strategy and send the classical random index, alongside the output of the first measurement, to {\bob}, who can then synchronize his local operations with those of {\alice}.

Let us now consider a complete set of linearly independent states $(\rho_\beta)_\beta$ spanning $\set{L}(\sH_C)$ and a family of informationally complete POVMs $(P_\alpha^{(j)})_{\alpha,j}$ spanning $\set{L}(\sH_{D_j})$ for all $j$. Eq.~(\ref{eq:condition}) can then be written as
\begin{equation*}
\exists \op{T}:\Tr{[\op{T}\mI]^{(j)}_y(\rho_\beta)\ P^{(j)}_\alpha}=\Tr{\mJ^{(j)}_y(\rho_\beta)\ P^{(j)}_\alpha},\quad\forall j,y,\alpha,\beta\;.
\end{equation*}

Since $\op{T}$ is chosen from a convex set, we can apply the separation theorem as follows:
\[
\forall\lambda_{jy\alpha\beta}\in\mathbb{R},\quad \max_{\op{T}}\sum_{j,y,\alpha,\beta}\lambda_{jy\alpha\beta}\Tr{[\op{T}\mI]^{(j)}_y(\rho_\beta)\ P^{(j)}_\alpha}\ge \sum_{j,y,\alpha,\beta}\lambda_{jy\alpha\beta}\Tr{\mJ^{(j)}_y(\rho_\beta)\ P^{(j)}_\alpha}\;.
\]
Introducing the Choi operators $C^{(j)}_y:=([\op{T}\mI]^{(j)}_y\otimes\id)\Omega$ and $D^{(j)}_y:=(\mJ^{(j)}_y\otimes\id)\Omega$, with $\Omega:=\sum_{m,n=1}^{d_C}\ket{m}\bra{n}\otimes\ket{m}\bra{n}$, we rewrite the above condition as
\[
\forall\lambda_{jy\alpha\beta}\in\mathbb{R},\quad \max_{\op{T}}\sum_{j,y,\alpha,\beta}\lambda_{jy\alpha\beta}\Tr{C^{(j)}_y\ P^{(j)}_\alpha\otimes \rho_\beta^\top}\ge \sum_{j,y,\alpha,\beta}\lambda_{jy\alpha\beta}\Tr{D^{(j)}_y\ P^{(j)}_\alpha\otimes \rho_\beta^\top}\;,
\]
where the superscript $\top$ denote the transposition done with respect to the basis $\{\ket{m}\}_m$ used to define the operator $\Omega$.

Absorbing the coefficients $\lambda_{jy\alpha\beta}$ in the operators $P^{(j)}_\alpha\otimes \rho_\beta^\top$, and introducing the operators $Z_{j,y}:=\sum_{\alpha,\beta}\lambda_{jy\alpha\beta}P^{(j)}_\alpha\otimes \rho_\beta^\top$, the condition~\eqref{eq:condition} can be equivalently reformulated as follows:
\begin{align}
&\forall \{Z_{j,y}:\text{self-adjoint}\},\nonumber\\
&\qquad\max_{\op{T}}\sum_{j,y}\Tr{C^{(j)}_y\ Z_{j,y}}\ge \sum_{j,y}\Tr{D^{(j)}_y\  Z_{j,y}}\;.\label{eq:to-be-rescaled}
\end{align}
(We recall that the dependence on $\op{T}$ is hidden in the operators $C^{(j)}_y$.)

Since by construction of the Choi operators
\[
\sum_y\PTr{1}{C^{(j)}_y}=\sum_y\PTr{1}{D^{(j)}_y}=\openone\;,\forall j\in\set{J}\;,
\]
it is possible to transform the operators $Z_{j,y}$ as follows
\[
Z_{j,y}\mapsto\frac{1}{K}\left(Z_{j,y}+\openone\otimes X_j+c\openone\otimes\openone\right)\;,
\]
for any choice of real constants $K$ and $c$, and self-adjoint operators $X_j$, without changing the inequality in Eq.~\eqref{eq:to-be-rescaled}. It is thus possible to restrict the comparison to families of operators $Z_{j,y}\ge 0$ that moreover satisfy $\sum_y\PTr{1}{Z_{j,y}}=\openone$, that is, operators that are themselves the Choi operators corresponding to programmable instruments. Denoting $Z_{j,y}\equiv \widetilde{Z}^{(j)}_y$,
\begin{align*}
&\forall \{\widetilde{Z}^{(j)}_y:\text{Choi operators of a programmable instrument}\},\\
&\qquad\max_{\op{T}}\sum_{j,y}\Tr{C^{(j)}_y\ \widetilde{Z}^{(j)}_y}\ge \sum_{j,y}\Tr{D^{(j)}_y\  \widetilde{Z}^{(j)}_y}\;.
\end{align*}

Further, since $\Tr{MN}=d\Tr{M\otimes N^\top\ \Phi^+}$, where $\Phi^+:=d^{-1}\Omega$, and since $\widetilde{Z}^{(j)}_y$ is a programmable instrument if and only if $(\widetilde{Z}^{(j)}_y)^\top$ is, we arrive at
\begin{align}\label{eq:condition2}
	&\forall\textrm{ programmable instruments } \{\widetilde{Z}^{(j)}_y\} \;,\nonumber\\ &\qquad \max_{\op{T}}\sum_{j,y}\Tr{(\widetilde{Z}^{(j)}_y\otimes C^{(j)}_y) \ (\Phi^+_{C'C}\otimes\Phi^+_{D_jD'_j})}\ge \sum_{j,y}\Tr{(\widetilde{Z}^{(j)}_y\otimes D^{(j)}_y)\ (\Phi^+_{C'C}\otimes\Phi^+_{D_jD'_j})}\;,
\end{align}
which, going back from Choi operators to the corresponding channels, reads
\begin{align}\label{eq:condition22}
	&\forall\textrm{ programmable instruments } (\mK^{(j)}_y)_{y,j} \;,\nonumber\\ &\qquad \max_{\op{T}}\sum_{j,y}\bra{\Phi^+_{D_j'D_j}}(\mK^{(j)}_y\otimes[\op{T}\mI]^{(j)}_y)(\Phi^+_{C'C})\ket{\Phi^+_{D_j'D_j}}\nonumber\\
 &\qquad\qquad\qquad\ge \sum_{j,y}\bra{\Phi^+_{D_j'D_j}}(\mK^{(j)}_y\otimes\mJ^{(j)}_y)(\Phi^+_{C'C})\ket{\Phi^+_{D_j'D_j}}\;.
\end{align}
Finally, in order to make formulas more symmetric, we notice that the above is also equivalent to the following:
\begin{align}\label{eq:condition3}
&\forall\textrm{ programmable instruments } (\mK^{(j)}_y)_{y,j} \;,\nonumber\\ &\qquad \max_{\op{T}}\sum_{j,y}\bra{\Phi^+_{D_j'D_j}}(\mK^{(j)}_y\otimes[\op{T}\mI]^{(j)}_y)(\Phi^+_{C'C})\ket{\Phi^+_{D_j'D_j}}\nonumber\\
&\qquad\qquad\ge \max_{\op{S}}\sum_{j,y}\bra{\Phi^+_{D_j'D_j}}(\mK^{(j)}_y\otimes[\op{S}\mJ]^{(j)}_y)(\Phi^+_{C'C})\ket{\Phi^+_{D_j'D_j}}\;.
\end{align}

\begin{figure}
    \centering
    \includegraphics[width=0.8\textwidth]{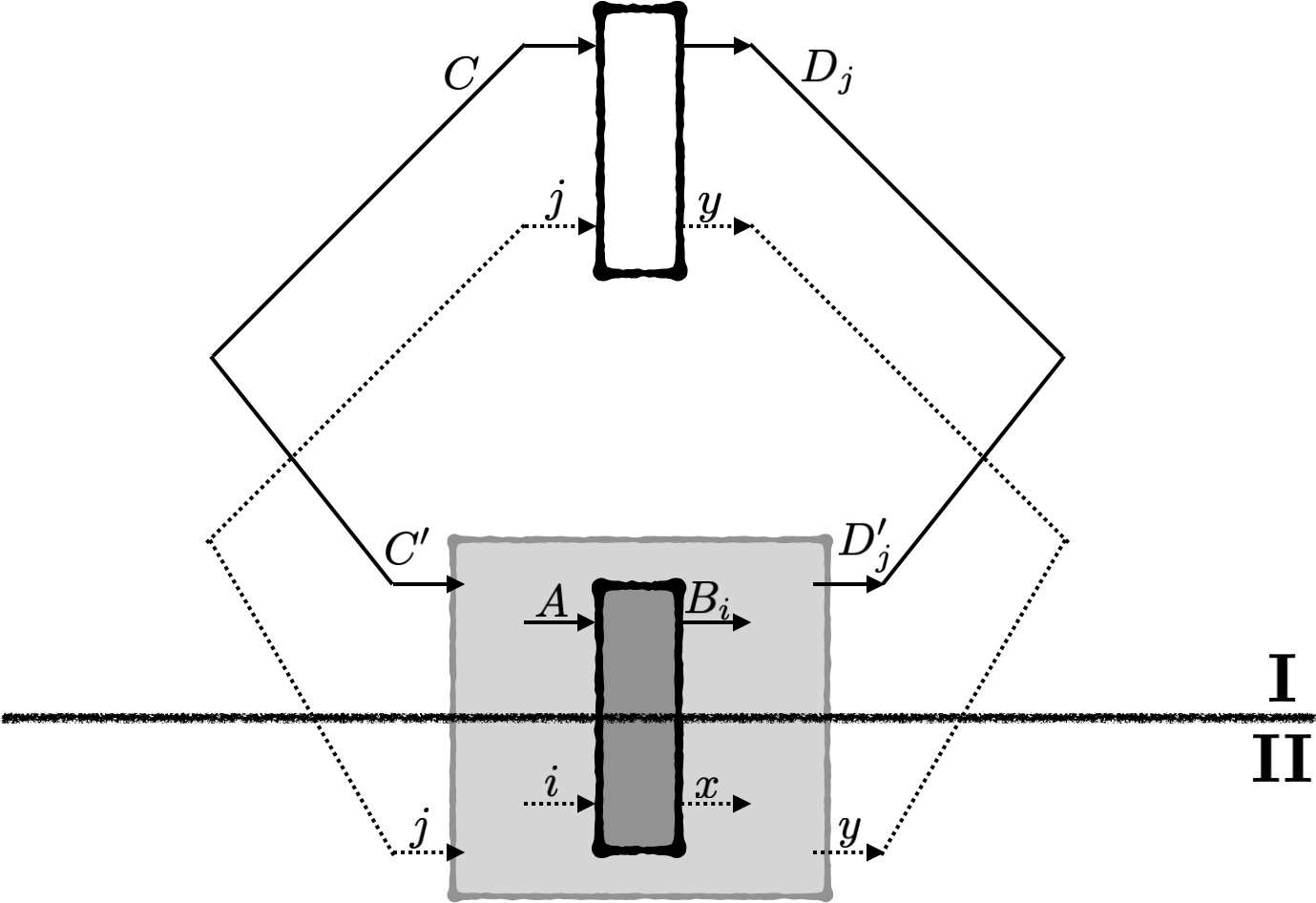}
    \caption{A \textbf{distributed classical-quantum guessing game}. The dark gray inner box is the programmable instrument initially shared by the two separated players. The referee's programmable instrument, with respect to which the game is defined, is the white box on top. The referee chooses a program value and send it to {\bob}. The referee also prepares a maximally entangled state, half of which is fed into the reference device, while the other half is handed over to {\alice}. The goal for the players is to guess the classical outcome of the referee's device, and, conditional on a correct guess, maximize the entanglement of {\alice}'s and the referee's quantum outputs.}
    \label{fig:synchro-game}
\end{figure}

We summarize the above construction in one definition and one theorem as follows.
\begin{definition}[distributed classical-quantum guessing games]
Two spatially separated players, {\alice} and {\bob}, initially share a programmable instrument $(\mI^{(i)}_x:A\to B_i)_{x\in\set{X},i\in\set{I}}$. A referee chooses a reference programmable instrument $(\mK^{(j)}_y:C\to D_j)_{y\in\set{Y},j\in\set{J}}$. In each round, the referee picks a program value at random from the set $\set{J}$ and sends it to {\bob}. At the same time, the referee prepares a maximally entangled state $\Phi^+_{CC'}$ and sends the $C'$ system to {\alice}. For each operational framework $\set{T}_{c}$, $\set{T}_q$, or $\set{T}_{ex}$, the expected utility associated to $(\mI^{(i)}_x)_{x,i}$ is computed as
\[
u_{\bullet}((\mI^{(i)}_x);(\mK^{(j)}_y)):=\max_{\op{T}\in\set{T}_{\bullet}}\sum_{j,y}\bra{\Phi^+_{D_j'D_j}}(\mK^{(j)}_y\otimes[\op{T}\mI]^{(j)}_y)(\Phi^+_{C'C})\ket{\Phi^+_{D_j'D_j}}\;,
\]
where $\bullet\in\{c,q,ex\}$. The situation is depicted in Fig.~\ref{fig:synchro-game}.
\end{definition}

By construction, each distributed classical-quantum guessing game defines a monotone $u_{\bullet}((\mI^{(i)}_x);(\mK^{(j)}_y))$, in the sense that
\[
u_{\bullet}((\mI^{(i)}_x);(\mK^{(j)}_y))\ge u_{\bullet}(([\op{T}\mI]^{(k)}_z);(\mK^{(j)}_y))\;,
\]
for any $\op{T}\in\set{T}_{\bullet}$. Using these monotones, it is possible to consider the following game-theoretic preorders:

\begin{definition}[statistical preorders]
Given two programmable instruments $(\mI^{(i)}_x:A\to B_i)_{x\in\set{X},i\in\set{I}}$ and $(\mJ^{(j)}_y:C\to D_j)_{y\in\set{Y},j\in\set{J}}$, we write
\[
(\mI^{(i)}_x)_{x,i}\supseteq_\bullet(\mJ^{(j)}_y)_{y,j}
\]
where $\bullet\in\{c,q,ex\}$, whenever
\[
u_{\bullet}((\mI^{(i)}_x);(\mK^{(j)}_y))\ge u_{\bullet}((\mJ^{(j)}_y);(\mK^{(j)}_y))\;,
\]
for all distributed classical-quantum guessing games $(\mK^{(j)}_y:C\to D_j)_{y,j}$.
\end{definition}

The main result of the above calculation is that the majorization preorders and the operational preorders are in fact equivalent.

\begin{theorem}[quantum Blackwell theorem for programmable instruments]
    The equivalence holds:
    \[
    (\mI^{(i)}_x)_{x,i}\supseteq_\bullet(\mJ^{(j)}_y)_{y,j}\iff (\mI^{(i)}_x)_{x,i}\succeq_\bullet(\mJ^{(j)}_y)_{y,j}\;,
    \]
   where $\bullet\in\{c,q,ex\}$. 
\end{theorem}

\begin{corollary}
    All classically compatible programmable instruments achieve the same expected utility in all classical-quantum guessing games. We denote this value as
    \[
        u^\star_{c}((\mK^{(j)}_y))\;.
    \]
    The same holds also for q-compatible and non-exclusive programmable instruments, and the corresponding threshold values are denotes $u^\star_{q}((\mK^{(j)}_y))$ and $u^\star_{ex}((\mK^{(j)}_y))$.
\end{corollary}

\begin{corollary}
    A programmable instrument $(\mI^{(i)}_x)_{x,i}$ is classically incompatible if and only if there exists a classical-quantum guessing game $(\mK^{(j)}_y)_{y,j}$ such that
    \[
    u_{c}((\mI^{(i)}_x);(\mK^{(j)}_y))> u^\star_{c}((\mK^{(j)}_y))\;.
    \]
    Analogous relations hold for q-incompatible and exclusive programmable instruments. In particular,
    \[
    u^\star_{c}((\mK^{(j)}_y))<u^\star_{q}((\mK^{(j)}_y))<u^\star_{ex}((\mK^{(j)}_y))\;.
    \]
\end{corollary}

\subsection{Constraining communication by timing: an example}

The resource theories of incompatibility that we considered in this work crucially depend on the way in which communication is restricted between {\alice} and {\bob}. Such restrictions hence need to be included also in the game-theoretic scenario of classical-quantum guessing games, lest the whole construction trivializes.

\begin{figure}
        \centering
        \includegraphics[width=\textwidth,keepaspectratio]{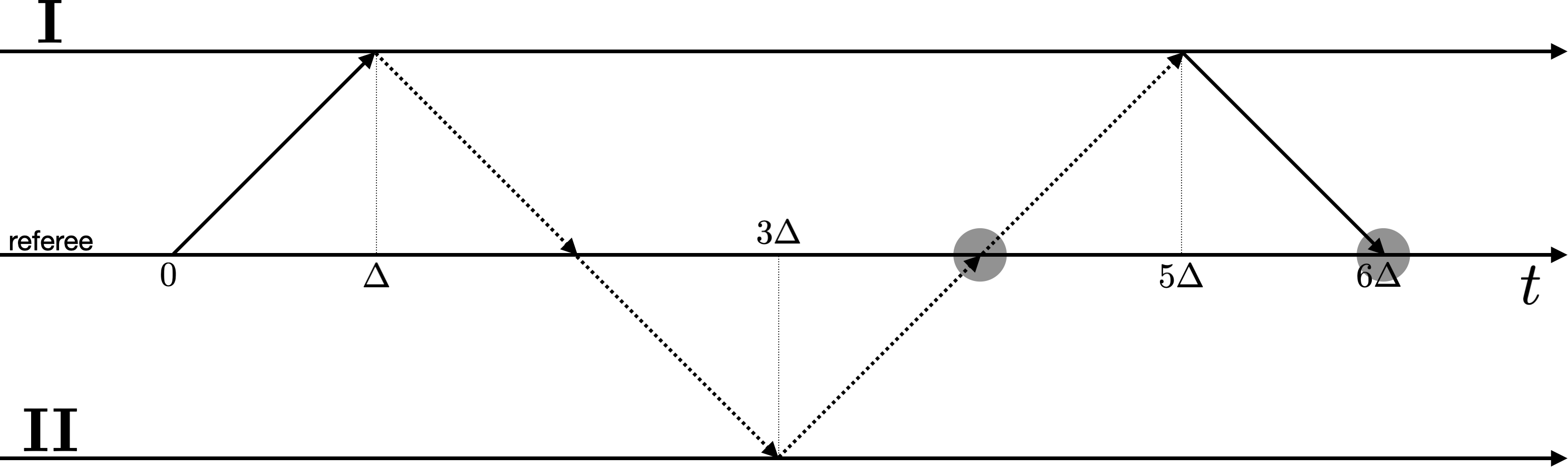}
        \caption{Example of a timed protocol that by construction allows two classically communicating players to simulate any q-compatible programmable instrument, but nothing beyond that. The small grey circles highlight the referee's local times at which she will receive the classical outcome ($t=4\Delta$) and quantum output ($t=6\Delta$).}
        \label{fig:timing-q-comp}
\end{figure}

One possibility is to constrain communication by timing, as routinely done, for example, in Bell tests~\cite{brunner2013Bell}. Fig.~\ref{fig:timing-q-comp} depicts a simple scenario in which, by the construction, two classically communicating players can implement any superoperation in $\set{T}_q$, but nothing beyond that. The protocol is constructed as follows:
\begin{enumerate}
\item the referee, sitting in between {\alice} and {\bob}, first sends the quantum state to {\alice}, who receives it at time $t=\Delta$ (we assume that all communications, classical and quantum, happen at constant speed, e.g., the speed of light);
\item {\alice} does something (local operations are accounted for as instantaneous) on the state she received and send some classical information to {\bob};
\item at time $t=2\Delta$ the referee sends the classical input to {\bob};
\item {\bob} receives both the referee's question and {\alice}'s information at time $t=3\Delta$, and uses them to process the classical output, which is immediately send back to the referee; at the same time, {\bob} sends classical information back to {\alice};
\item the referee receives the classical outcome from {\bob} at time $t=4\Delta$;
\item finally, {\alice} receives the classical feedback from {\bob} at time $t=5\Delta$, uses that to post-process the quantum output, and sends this back to the referee, who will receive it at time $t=6\Delta$.
\end{enumerate}

Clearly, if {\alice} and {\bob} share a quantum noiseless channel, {\alice}  could immediately send the quantum state she received from the referee to {\bob}, who could locally apply any instrument of his choosing. He would then send the classical outcome to the referee and the quantum output back to {\alice}, who could return it to the referee, as requested, at time $t=6\Delta$. In other words, the timing above allows two quantum communicating parties to produce any programmable instruments, even incompatible ones. This is in agreement with our viewpoint, according to which quantum communication\footnote{Direct quantum communication or indirect quantum communication, via quantum teleportation. The latter would anyway require the players to have access to a noiseless quantum memory (i.e., noiseless quantum communication in time) to store the entanglement until required.} constitutes a resource that can be used to ``generate'' q-incompatibility.

It is important to stress however, that the above is by no means the only timing strategy possible. For example, an alternative strategy could grant {\bob} some extra time, requiring him to return the classical outcome to the referee not immediately, but so that it reaches the referee at time $t=6\Delta$, that is, together with {\alice}'s quantum output. This alternative picture could be useful to permit the two players to make use of some pre-shared bipartite channel as a side-resource, for example, a pre-shared programmable instrument, which could provide them with an advantage, even if the communication between them always remains classical.

The role of timing in classical-quantum guessing games, and how it constrains communication in relation to quantum incompatibility, seems to constitute an interesting problem that we leave open for future investigations.





\section*{Acknowledgments}

The authors are grateful to Teiko Heinosaari and Masanao Ozawa for insightful comments. F.~B. acknowledges support from MEXT Quantum Leap Flagship Program (MEXT QLEAP) Grant No. JPMXS0120319794; from MEXT-JSPS Grant-in-Aid for Transformative Research Areas (A) ``Extreme Universe'', No. 21H05183; from JSPS KAKENHI Grants No. 20K03746 and No. 23K03230. S.~M. acknowledges support from the ``Nagoya University Interdisciplinary Frontier Fellowship'' supported by Nagoya University and JST, for the establishment of university fellowships towards the creation of science technology innovation, Grant Number JPMJFS2120. P.~P. and A.~T. acknowledge the financial support of Elvia and Federico Faggin Foundation (Silicon Valley Community Foundation Project ID\#2020-214365). P.~P.~acknowledges financial support from PNRR MUR project PE0000023.

\bibliographystyle{alphaurl}
\bibliography{myref,library}

\end{document}